\documentclass[aps,prb,twocolumn,showpacs]{revtex4}
\usepackage{bm}
\usepackage{graphicx}

\newcommand{\vrho}{\bm{\rho}}
\newcommand{\hz}{\hat{\bm{z}}}
\newcommand{\hr}{\hat{\bm{\rho}}}
\newcommand{\htheta}{\hat{\bm{\theta}}}

\begin{document} 
\title{Inductive measurements of third-harmonic voltage and critical current density in bulk superconductors}
\author{Yasunori Mawatari, Hirofumi Yamasaki, and Yoshihiko Nakagawa}
\affiliation{Energy Electronics Institute,
   National Institute of Advanced Industrial Science and Technology, \\
   Tsukuba, Ibaraki 305--8568, Japan}
\date{\today}

\begin{abstract}
We propose an inductive method to measure critical current density $J_c$ in bulk superconductors. 
In this method, an ac magnetic field is generated by a drive current $I_0$ flowing in a small coil mounted just above the flat surface of superconductors, and the third-harmonic voltage $V_3$ induced in the coil is detected. 
We present theoretical calculation based on the critical state model for the ac response of bulk superconductors, and we show that the third-harmonic voltage detected in the inductive measurements is expressed as $V_3= G_3\omega I_0^2/J_c$, where $\omega/2\pi$ is the frequency of the drive current, and $G_3$ is a factor determined by the configuration of the coil.
We measured the $I_0$-$V_3$ curves of a melt-textured $\rm YBa_2Cu_3O_{7-\delta}$ bulk sample, and evaluated the $J_c$ by using the theoretical results. 
\end{abstract}
\maketitle
   \thispagestyle{myheadings}\markright{\today} 

Large melt-textured RE-Ba-Cu-O (where RE represents rare-earth elements) superconductors can trap strong magnetic fields, and have been developed for high-field engineering applications.~\cite{Tomita03} 
The trapped field is proportional to the critical current density $J_c$, and a nondestructive and contactless method is desirable to measure $J_c$ distribution in large-sized superconductors. 

Claassen {\it et al}.~\cite{Claassen91} proposed a nondestructive and contactless method to measure $J_c$ in superconducting {\em films}, and the scientific basis of this inductive technique was theoretically justified.~\cite{Poulin93,Mawatari02} 
In that method an ac magnetic field is generated by a sinusoidal drive current $I_0\cos\omega t$ flowing in a coil close to the upper surface of a superconducting film, and the amplitude $V_3$ of the third-harmonic voltage $V_3\cos(3\omega t+\vartheta_3)$ induced in the coil is simultaneously measured. 
For superconducting films of thickness $d_f$, $V_3$ is detected when $I_0>I_{c0}$, where $I_{c0}\propto J_cd_f$ is the threshold current at which the sheet current of the film reaches its critical value $J_cd_f$ and the ac magnetic field penetrates below the film.~\cite{Mawatari02} 
The $J_c$ is determined from $I_{c0}$,~\cite{Claassen91} and the current-voltage characteristics can be determined by measuring the frequency $\omega/2\pi$ dependence of $J_c$.~\cite{Yamasaki03} 

In this letter, we theoretically calculate the ac response of {\em bulk superconductors} based on the critical state model,~\cite{Bean62,Bean64} and derive the relationship between $I_0$ and $V_3$ that are measured in the inductive method. 
Experimental setups and measured quantities in the inductive method for {\em bulk superconductors} proposed in this letter are basically the same as those in the previously developed inductive method for {\em films}.~\cite{Claassen91} 
Because the thickness of bulk superconductors is much larger than that for superconducting films, an ac magnetic field generated by a small coil placed above the bulk superconductors cannot penetrate below the lower surface of the superconductors. 
The harmonic voltages of bulk superconductors, however, arise from irreversible vortex motion,~\cite{Bean64} even when the ac magnetic field is shielded below the lower surface of the bulk superconductor.
Our theoretical calculations described in this letter show that the $I_0$-$V_3$ curves for bulk superconductors significantly differ from those for films,~\cite{Claassen91,Poulin93,Mawatari02,Yamasaki03} but are similar to those for superconducting rods.~\cite{Bean64,Love66}

First, we theoretically calculate the ac response of a bulk superconductor of thickness $d$ situated at $-d<z<0$, where the upper surface is at $z=0$ and the lower surface at $z=-d$. 
A single coil mounted just above the superconductor is used to generate an ac magnetic field and to detect the induced voltage.~\cite{Claassen91} 
The bulk superconductor is considered to extend infinitely in the $xy$ plane, and thus the electromagnetic fields are not affected by the edges of the bulk superconductor sample. 
The edge effects are negligible if the distance between the sample edge and the coil edge is larger than the radius of the coil.~\cite{Claassen91,Hochmuth94} 
We consider a coil of circular shape that produces axially symmetric fields independent of azimuth $\theta$ in cylindrical coordinates $(\bm{\rho},z)= (\rho,\theta,z)$. 
Current density $\bm{J}=J(\rho,z,t)\htheta$, sheet current $\bm{K}=K(\rho,t)\htheta$, and vector potential $\bm{A}=A(\rho,z,t)\htheta$ have only azimuthal components. 
Although the magnetic field $\bm{H}$ has both a radial component $H_{\rho}$ parallel to the superconductor surface and a $z$ component $H_z$ normal to the surface, we assume that $H_{\rho}$ plays a crucial role in determining the ac response of the bulk superconductors. 
The shielding current $\bm{J}_s$ flowing near the surface prevents penetration of magnetic flux into the bulk superconductors, and $\bm{J}_s$ mostly cancels out $H_z$ near the surface of the bulk superconductors, $|H_{\rho}|\gg|H_z|$ for $z\approx 0$. 

A drive current $I_0\cos\omega t$ flowing in the coil produces an ac magnetic field, and its radial component at $z=0$ is given by
\begin{equation}
   H_0(\rho,t)=-I_0F_1(\rho)\cos\omega t .
\label{H0_I0F1}
\end{equation}
The coil-factor function $F_1(\rho)$ is determined by the configuration of the coil,
\begin{eqnarray}
   F_1(\rho) &=& \frac{N}{4\pi S_c}
      \int_{R_1}^{R_2}d\rho' \int_0^{2\pi}d\theta \int_{Z_1}^{Z_2}dz \,
      \rho' z\cos\theta
\nonumber\\
   && \times (z^2+\rho^2+\rho'^2-2\rho\rho'\cos\theta)^{-3/2} ,
\label{F-coil-factor}
\end{eqnarray}
where $N$ is the number of windings, $S_c=(R_2-R_1)(Z_2-Z_1)$ is the cross-sectional area of the coil in the $\rho z$ plane at $R_1<\rho<R_2$ and $Z_1<z<Z_2$ where $Z_1>0$. 

Responding to the ac magnetic field $H_0$, shielding current density $\bm{J}_s$ is induced near the surface of the bulk superconductor, $-\Lambda_0<z<0$, where $\Lambda_0$ is the flux-penetration depth that is smaller than the thickness $d$ of the bulk superconductor. 
The vector potential arising from $\bm{J}_s$ is given by 
\begin{eqnarray}
   \lefteqn{\bm{A}_s(\vrho,z,t)} 
\nonumber\\
   && =\frac{\mu_0}{4\pi} \int_{S_{sc}}d^2\vrho' \int_{-\Lambda_0}^0dz'\, 
      \frac{\bm{J}_s(\vrho',z',t)}{%
      \left(|\vrho-\vrho'|^2 +|z-z'|^2\right)^{1/2}} , 
\label{As-Js}
\end{eqnarray}
where the integral region $S_{sc}$ corresponds to the surface of the bulk superconductor in the $xy$ plane, i.e., $0<\rho'<\infty$ and $0<\theta'<2\pi$. 
The $\bm{A}_s$ is expressed as the multipole expansion, $\bm{A}_s=\bm{A}_{s1}+\bm{A}_{s2}+\cdots$ for $|z'|\ll R_2$, where 
\begin{eqnarray}
   \bm{A}_{s1}(\vrho,z,t) &=& \frac{\mu_0}{4\pi} \int_{S_{sc}}d^2\vrho'
      \frac{\bm{K}_s(\vrho',t)}{%
      \left(|\vrho-\vrho'|^2 +z^2\right)^{1/2}} ,
\label{As1-Ks}\\
   \bm{A}_{s2}(\vrho,z,t) &=& \frac{\mu_0}{4\pi} \int_{S_{sc}}d^2\vrho'
      \frac{\bm{M}_s(\vrho',t)\times(z\hz)}{%
      \left(|\vrho-\vrho'|^2+z^2\right)^{3/2}} .
\label{As2-Ms}
\end{eqnarray}
The $\bm{K}_s(\vrho,t)=\int_{-\Lambda_0}^0dz \bm{J}_s(\vrho,z,t)$ is the sheet current, and $\bm{M}_s(\vrho,t) =M_s(\rho,t)\hr =\int_{-\Lambda_0}^0dz (z\hz)\times\bm{J}_s(\vrho,z,t)$ is the sheet magnetic moment. 

The magnetic flux linked in the coil of $N\gg 1$ wound by a thin wire is calculated as 
\begin{eqnarray}
   \Phi(t) &=& \frac{2\pi N}{S_c}\int_{R_1}^{R_2}\rho d\rho
      \int_{Z_1}^{Z_2}dz A(\rho,z,t)
\nonumber\\
   &=& L_cI_0\cos\omega t +\Phi_{s1}(t)+\Phi_{s2}(t)+\cdots ,
\label{flux_d+s}
\end{eqnarray}
where the first term in the right-hand side of Eq.~(\ref{flux_d+s}) arises from the drive current, and $L_c$ is the self-inductance of the coil. 
The remaining terms in Eq.~(\ref{flux_d+s}), $\Phi_{s1}+\Phi_{s2}+\cdots$, arise from the shielding current in the superconductor; $\Phi_{s1}$ is due to $K_s$ and $\Phi_{s2}$ is due to $M_s$. 
The ratio of $|\Phi_{s2}|/|\Phi_{s1}|$ is on the order of $\Lambda_0/R_2$ for $Z_2\lesssim R_2$, whereas $|\Phi_{s1}|$ is on the order of $L_cI_0$. 
Because the ratio $\Lambda_0/R_2$ is typically on the order of $0.01$ or less, we can neglect $\Phi_{s2}$ to obtain the linear response to $I_0$. 
The nonlinear response of bulk superconductors, however, arises from $\Phi_{s2}$, as we show below.~\cite{V3_film} 

When the ac field does not reach the lower surface of a bulk superconductor (i.e., $2|H_0|<J_cd$ and $\Lambda_0<d$), the sheet current is given by $K_s=2H_0$ (where $|K_s|<J_cd$), the ac magnetic field at the surface of the superconductor ($z=0$) is also given by $2H_0$,~\cite{Mawatari02} 
and the magnetic flux due to the sheet current is simply expressed as $\Phi_{s1}=-L_{ci}I_0\cos\omega t$, where $L_{ci}$ is the mutual inductance between the coil and its image.~\cite{Claassen91} 
The $\Phi_{s2}$ is calculated as the line integral of Eq.~(\ref{As2-Ms}) along the wire of the coil, and is given by $\Phi_{s2}(t)= -2\pi\mu_0\int_0^{\infty}d\rho\,\rho F_1(\rho)M_s(\rho,t)$, where $F_1(\rho)$ is given by Eq.~(\ref{F-coil-factor}). 
Induced voltage in the coil is given by $V(t)= R_cI_0\cos\omega t-d\Phi/dt$, which can be rewritten as 
\begin{eqnarray}
   V(t) &=& R_cI_0\cos\omega t +\omega(L_c-L_{ci})I_0\sin\omega t
\nonumber\\
   && \mbox{}+2\pi\mu_0 \int_0^{\infty}d\rho\,\rho F_1(\rho)
      \frac{\partial M_s(\rho,t)}{\partial t} ,
\label{V-t}
\end{eqnarray}
where $R_c$ is the resistance of the coil. 
The amplitude of the last term of the right-hand side of Eq.~(\ref{V-t}) is much smaller than the other terms, but the first and second terms do not contribute to the harmonic response. 

The shielding current $\bm{J}_s=J_s(\rho,t)\htheta$ induced by the ac magnetic field of Eq.~(\ref{H0_I0F1}) in the critical state model~\cite{Bean62,Bean64} is given as follows: 
$J_s(\rho,t)=\mp J_c$ for $-\Lambda_{\mp}(\rho,t)<z<0$, and $J_s(\rho,t)=\pm J_c$ for $-\Lambda_0(\rho)<z<-\Lambda_{\mp}(\rho,t)$, where $J_c$ is the critical current density. 
The signs `$\pm$' (`$\mp$') represents `$+$' (`$-$') for $0<\omega t<\pi$ and `$-$' (`$+$') for $\pi<\omega t<2\pi$.~\cite{even-harmonics} 
The flux-penetration depths are given by $\Lambda_0(\rho)= 2I_0F_1(\rho)/J_c$ and $\Lambda_{\pm}(\rho,t)=\Lambda_0(\rho)(1\mp\cos\omega t)/2$. 
Then, the sheet magnetic moment $M_s(\rho,t)= \int_{-\Lambda_0}^0dz (-z)J_s(\rho,z,t)$ is calculated as 
\begin{eqnarray}
   M_s(\rho,t) &=& \frac{1}{2} J_c\Lambda_0(\rho)^2
      \left( \cos\omega t\pm\frac{1}{2}\sin^2\omega t \right) ,
\nonumber\\
   &=& \left[2I_0^2F_1(\rho)^2/J_c\right]
\nonumber\\
   && \times\left( \cos\omega t -\frac{4}{\pi}\sum_{k=1}^{\infty}
      \frac{\sin(2k-1)\omega t}{(2k-3)(4k^2-1)} \right) . 
\label{Ms-rt-Fourier}
\end{eqnarray}
Substitution of Eq.~(\ref{Ms-rt-Fourier}) into Eq.~(\ref{V-t}) yields
\begin{eqnarray}
   V(t) &=& V_1\cos(\omega t+\vartheta_1)
      -\sum_{n=2}^{\infty} V_n \cos n\omega t .
\label{V-Fourier}
\end{eqnarray}
The harmonic voltages are given by $V_n=0$ for even $n=2,\,4,\,6,...$~\cite{even-harmonics} and $V_n=5V_3/(n^2-4)$ for odd $n=3,\,5,\,7,...$. 
The third-harmonic voltage is
\begin{eqnarray}
   V_3 &=& G_3\,\omega I_0^2/J_c ,
\label{V3-I0_bulk}\\
   G_3 &=& \frac{16\mu_0}{5}
      \int_0^{\infty}d\rho\,\rho F_1^{\,3}(\rho) ,
\label{G3_coil-factor}
\end{eqnarray}
where $G_3$ is the coil factor determined by the configuration of the coil. 
The fundamental voltage $V_1\cos(\omega t+\vartheta_1)$ in Eq.~(\ref{V-Fourier}) is affected by the impedance of the coil, $R_c$ and $\omega(L_c-L_{ci})$. 
Equation~(\ref{V3-I0_bulk}) is similar to $V_3$ derived by Bean~\cite{Bean64} to determine {\em averaged} $J_c$ over the surface of {\em cylindrical superconductors} exposed to a {\em uniform} ac magnetic field. 
Note that Eq.~(\ref{V3-I0_bulk}) is applicable to evaluate {\em local} $J_c$ distribution at the flat surface of {\em large bulk superconductors} responding to a {\em nonuniform} ac magnetic field.

Next, we demonstrate evaluation of $J_c$ by using the theoretical result of Eq.~(\ref{V3-I0_bulk}) from our experimental data for a melt-textured $\rm YBa_2Cu_3O_{7-\delta}$ (YBCO) bulk superconductor sample. 
The YBCO sample was $12.1\times 12.7$mm$^2$ and $0.9\,$mm thick. 
The third-harmonic voltage induced by the YBCO sample was measured by using the experimental setup described in Ref.~\onlinecite{Yamasaki03}. 
Although our experimental setup was similar to that in Refs.~\onlinecite{Claassen91} and \onlinecite{Poulin93}, an additional cancel coil was needed to compensate for harmonic noise voltages.~\cite{Yamasaki03} 
The coil used to generate the ac magnetic field and to detect the induced voltage had $N=400$ turns and dimensions of inner diameter of $2R_1=1.0\,$mm, outer diameter of $2R_2=3.5\,$mm, and height of $Z_2-Z_1=1.0\,$mm.
The distance between the coil and the surface of the sample was $Z_1=0.25\,$mm.
A dc magnetic field $B_{\rm dc}$ was applied perpendicular to the flat surface of the sample (parallel to the $c$ axis of YBCO). 

\begin{figure}[t]
\includegraphics{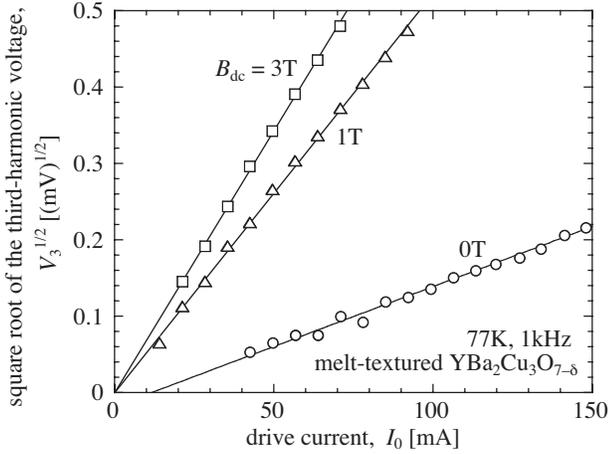}
\caption{%
Drive current $I_0$ dependence of the square root of the third-harmonic voltage $V_3^{1/2}$ for a melt-textured $\rm YBa_2Cu_3O_{7-\delta}$ sample at $77\,$K and $\omega/2\pi=1$\,kHz. 
A dc magnetic field $B_{\rm dc}$ was applied perpendicular to the wide surface of the sample. 
}\label{Fig_I0-V3}
\end{figure}
%

Figure \ref{Fig_I0-V3} shows the square root of the measured third-harmonic voltage $V_3^{1/2}$ as a function of the drive current $I_0$ for the YBCO bulk superconductor sample. 
The fitted lines for the data of $B_{\rm dc}=1$ and 3\,T intersect the origin, and are consistent with Eq.~(\ref{V3-I0_bulk}). 
We can evaluate $J_c$ from the slope of the lines in Fig.~\ref{Fig_I0-V3} by using Eq.~(\ref{V3-I0_bulk}), where the coil factor calculated by using Eqs.~(\ref{F-coil-factor}) and (\ref{G3_coil-factor}) is $G_3=735\,\Omega\cdot$sec/m$^2$ for the coil used in our measurements. 
The resulting $J_c$ values are $1.7\times 10^8$ and $1.0\times 10^8\,$A/m$^2$ for $B_{\rm dc}=1$ and 3\,T, respectively. 

The fitted line for the data of $B_{\rm dc}=0\,$T in Fig.~\ref{Fig_I0-V3} does not intersect the origin, suggesting the following empirical equation for $I_0$-$V_3$:
\begin{equation}
   V_3= G_3\,\omega (I_0-I_{c1})^2/J_c
   \quad\mbox{for } I_0>I_{c1} ,
\label{V3_mod}
\end{equation}
where $J_c=1.9\times 10^9\,$A/m$^2$.
The threshold current is $I_{c1}=11\,$mA, which corresponds to the ac magnetic field $2\mu_0|H_0|\sim 2\,$mT at the surface of the bulk superconductor sample. 
The threshold current $I_{c1}$ might be reflected by the lower critical field or surface barrier field.
The importance of such barrier fields has been realized in harmonic-voltage measurements.~\cite{Love66}

A rough estimate of the electric field $E_s$ induced at the surface of the bulk superconductor is given as the product of angular frequency $\omega$, ac magnetic induction $2\mu_0|H_0|$, and the field penetration depth $\Lambda_0\sim 2|H_0|/J_c$:
\begin{equation}
   E_s\sim 2\mu_0\omega |H_0|\Lambda_0 \sim 4\mu_0\omega |H_0|^2/J_c .
\label{E_estimation}
\end{equation}
Both the $E_s$ induced in the bulk superconductors and $V_3$ induced in the coil are proportional to $\omega I_0^2/J_c$, and thus, the ratio $V_3/E_s$ ($\sim 5\,$cm for the coil used in our measurements) is governed only by the configuration of the coil. 
The third-harmonic voltage of $V_3\sim 0.05\,$mV, which is the typical order of $V_3$ shown in Fig.~\ref{Fig_I0-V3}, implies that the electric field induced at the surface of the bulk superconductor is on the order of $E_s\sim 1\,$mV/m, which is much larger than that in the magnetization measurements.~\cite{Mawatari97} 
Because $E_s\propto\omega$, we can measure $J_c$-$E_s$ curves (i.e., current-voltage characteristics) by measuring $\omega$ dependence of $J_c$, as demonstrated for films.~\cite{Yamasaki03}

In summary, we theoretically and experimentally investigated the third-harmonic voltage $V_3$ in response to an ac magnetic field generated by a sinusoidal drive current $I_0$ in a small coil placed just above bulk superconductors. 
The theoretical third-harmonic voltage induced in the coil is given by Eq.~(\ref{V3-I0_bulk}). 
The experimental data of melt-textured YBCO bulk samples for $B_{\rm dc}\geq 1$\,T are consistent with Eq.~(\ref{V3-I0_bulk}), and the $J_c$ values can thus be evaluated from the experimental data. 
These results show that the distribution of $J_c$ in large-sized bulk superconductors can be determined by measuring $I_0$-$V_3$ curves with a scanning coil (or fixed multiple coils) and using Eq.~(\ref{V3-I0_bulk}). 
We can measure $J_c$ only {\em near the surface} (i.e., $J_c$ for $-\Lambda_0<z<0$) of bulk superconductors, whereas information on $J_c$ deep inside bulk superconductors can not be obtained by this method. 

This work was done as part of the Super-ACE project (a project for the R\&D of fundamental technologies for superconducting ac power equipment) of the Ministry of Economy, Trade, and Industry of Japan.


\begin{thebibliography}{99}
\bibitem{Tomita03}
M. Tomita and M. Murakami, Nature {\bf 421}, 517 (2003), 
and references therein.
\bibitem{Claassen91}
J. H. Claassen, M. E. Reeves, and R. J. Soulen, Jr., 
Rev. Sci. Instrum. {\bf 62}, 996 (1991).
\bibitem{Poulin93}
G. D. Poulin, J. S. Preston, and T. Strach, \prb {\bf 48}, 1077 (1993).
\bibitem{Mawatari02}
Y. Mawatari, H. Yamasaki, and Y. Nakagawa, \apl {\bf 81}, 2424 (2002).
\bibitem{Yamasaki03}
H. Yamasaki, Y. Mawatari, and Y. Nakagawa, \apl {\bf 82}, 3275 (2003).
\bibitem{Bean62}
C. P. Bean, \prl {\bf 8}, 250 (1962).
\bibitem{Bean64}
C. P. Bean, \rmp {\bf 36}, 31 (1964).
\bibitem{Love66}
G. R. Love, J. Appl. Phys. {\bf 37}, 3361 (1966);
H. A. Ullmaier, Phys. Stat. Sol. {\bf 17}, 631 (1966);
S. T. Sekula and J. H. Barrett, \apl {\bf 17}, 204 (1970);
D. M. Kroeger, C. C. Koch, and W. A. Coghlan, 
J. Appl. Phys. {\bf 44}, 2391 (1973);
D. M. Kroeger, C. C. Koch, and J. P. Charlesworth, 
J. Low Temp. Phys. {\bf 19}, 493 (1975).
\bibitem{Hochmuth94}
H. Hochmuth and M. Lorenz, Physica (Amsterdam) {\bf 220C}, 209 (1994);
ibid. {\bf 265C}, 335 (1996).

\bibitem{V3_film}
For superconducting films, nonlinear response arises from $\Phi_{s1}$ when $I_0>I_{c0}$, and thus the small contribution from $\Phi_{s2}$ can be neglected to obtain the behavior of harmonic voltages for films, as done in the calculation in Ref.~\onlinecite{Mawatari02}. 
When $I_0<I_{c0}$, on the other hand, linear response of $\Phi_{s1}$ does not affect the harmonic voltage, and thus $\Phi_{s2}$ plays a crucial role in harmonic voltages. 
The contribution to harmonic voltages from $\Phi_{s2}\propto 1/J_c$ is, however, too small to detect in experiments for films, because $J_c$ of films is typically 100 times larger than $J_c$ of bulk materials. 
Therefore, $V_3\approx 0$ for $I_0<I_{c0}$ in films.~\cite{Claassen91,Poulin93,Mawatari02,Yamasaki03} 
\bibitem{even-harmonics}
We assume the periodicity of $J_s(\rho,t)=-J_s(\rho,t+\pi/\omega)$, which leads to the disappearance of the even harmonics, $V_n=0$ for $n$ even. 
Such behavior of $J_s$ and $V_n$ is not accurate because $J_c$ for $0<\omega t<\pi$ is not identical to $J_c$ for $\pi<\omega t<2\pi$ when $B_{\rm dc}\neq 0$. 
This assumption of the periodicity of $J_s$, however, is a good approximation and the even harmonics are generally much smaller than the odd harmonics. 

\bibitem{Mawatari97}
Y. Mawatari, A. Sawa, H. Obara, M. Umeda, and H. Yamasaki, \apl {\bf 70}, 2300 (1997); H. Yamasaki and Y. Mawatari, Supercond. Sci. Technol. {\bf 13}, 202 (2000).

\end{thebibliography}
\end{document}